\documentclass[journal=jctcce,manuscript=article]{achemso}

\usepackage{amsmath} % Formula subscripts using \ch{}
\usepackage[T1]{fontenc} % Use modern font encodings
\usepackage{setspace,lipsum}
\author{Jayashrita Debnath}
\affiliation[]{Department of Chemistry and Applied Biosciences, ETH Zurich, c/o USI Campus, Via Giuseppe Buffi 13, CH-6900, Lugano, Switzerland}
\alsoaffiliation{Facolt\`{a} di Informatica, Istituto di Scienze Computazionali,  Universit\`{a} della Svizzera Italiana (USI), Via Giuseppe Buffi 13, CH-6900, Lugano, Switzerland}
\author{Michele Invernizzi}
\affiliation[]{Department of Physics, ETH Zurich, c/o USI Campus, Via Giuseppe Buffi 13, CH-6900, Lugano, Switzerland}
\alsoaffiliation{Facolt\`{a} di Informatica, Istituto di Scienze Computazionali,  Universit\`{a} della Svizzera Italiana (USI), Via Giuseppe Buffi 13, CH-6900, Lugano, Switzerland}
\author{Michele Parrinello}
\affiliation{Department of Chemistry and Applied Biosciences, ETH Zurich, c/o USI Campus, Via Giuseppe Buffi
 13, CH-6900, Lugano, Switzerland}
\alsoaffiliation{Facolt\`{a} di Informatica, Istituto di Scienze Computazionali,  Universit\`{a} della Svizzera Italiana (USI), Via Giuseppe Buffi 13, CH-6900, Lugano, Switzerland}
\alsoaffiliation{Istituto Italiano di Tecnologia, Via Morego 30, 16163 Genova, Italy}
\email{parrinello@phys.chem.ethz.ch.}

\title[Enhanced sampling of transition states]
  {Enhanced sampling of transition states}

\begin{document}

%\begin{tocentry}
%\begin{center}
%\includegraphics[scale=1]{TOC.pdf}
%\label{For Table of Contents Only}
%\end{center}
%\center{For Table of Contents Only}
%\end{tocentry}

\begin{abstract}
The free energy landscapes of several fundamental processes are characterized by high barriers separating long-lived metastable states. In order to explore these type of landscapes enhanced sampling methods are used. While many such methods are able to obtain sufficient sampling in order to draw the free energy, the transition states are often sparsely sampled. We propose an approach based on the Variationally Enhanced Sampling Method to enhance sampling in the transition region. To this effect, we introduce a dynamic target distribution which uses the derivative of the instantaneous free energy surface to locate the transition regions on the fly and modulate the probability of sampling different regions. Finally, we exemplify  the effectiveness of this approach in enriching the number of configurations in the transition state region in the cases of a chemical reaction and of a nucleation process.
\end{abstract}

\section{Introduction}
One of the crucial problems in the atomistic simulation of systems of chemical-physical relevance is the study of rare events. Typical rare events are chemical reactions, phase transitions, and biomolecule conformational changes.  These events take place on a time scale that cannot be reached by straightforward molecular dynamics (MD) and require enhanced sampling methods to be studied. A vast class of such methods
\cite{Torrie1977,Laio2002,Maragliano2006,Barducci2008,Darve2008,Valsson2014,Valsson2015} 
relies on the identification of an appropriate set  of collective  variables (CVs)  whose fluctuations are then suitably enhanced.  The CVs, here indicated as ${\boldsymbol{s}}={\boldsymbol{s}}({\boldsymbol{R}})$ ,  are functions of the atomic coordinates ${\boldsymbol{R}}$,  and are meant to encode the slow degrees of freedom of the system. In our group we have focused on two such approaches, namely Metadynamics \cite{Laio2002} and Variationally Enhanced Sampling (VES) \cite{Valsson2014}. One of the outputs of these calculations is the free energy $F({\boldsymbol{s}})$ as a function of the chosen CVs. With adroitly chosen CVs, the $F({\boldsymbol{s}})$ separates the metastable states and the apparent TS does coincide with the real one. Luckily much progress  has been made in finding \cite{Mendels2018} and ameliorating CVs \cite{McCarty2017a,Tiwary2015a,Sultan2017}
and  we shall assume here that we are in a situation such that the CVs pass close to the real transition state (TS).

However, even with good CVs, in a standard Metadynamics or VES calculation the transition state is not as well sampled as one would like it to be. In fact in these methods based on a dynamical approach in which the $F({\boldsymbol{s}})$ is progressively filled by a bias, most of the sampling effort is spent in exploring the $F({\boldsymbol{s}})$ minima and the transition states (TS) are sampled only when the system moves from one basin to another. This is an event that, while accelerated by the bias, is comparatively rare. Given the relevance that the transition state plays in understanding the process under study and eventually guiding it towards a desired end, it would be of great help to have a method that partitions sampling more evenly between the different stationary points of $F({\boldsymbol{s}})$ and harvests a much larger set of transition state configurations.  

In order to achieve this result we shall use one of the most attractive features of VES, namely the possibility of searching for a bias such that the  distribution in the biased ensemble is equal to a pre-assigned one. This leads to an enhanced sampling of the transition state region whose properties can then be studied in great detail.  We apply the method thus developed to the study of a bimolecular nucleophilic substitution (S$_\text{N}$2) reaction and to a liquid-vapor transition, which is a prototypical nucleation process.

\section{Method}
\subsection{Variationally Enhanced Sampling}
In VES, the bias potential $V( {\boldsymbol{s}} ) $ is obtained by optimizing the functional
\begin{equation}
\label{vesfunc}
\Omega[V({\boldsymbol{s}})] = \frac{1}{\beta} log \frac{\int d {\boldsymbol{s}} \ e^{ {- \beta (V({\boldsymbol{s}}) + F({\boldsymbol{s}}))}}}{\int d{\boldsymbol{s}} \ e^{ - \beta F({\boldsymbol{s}})} } + \int d{\boldsymbol{s}} \ p({\boldsymbol{s}}) V({\boldsymbol{s}}) \, ,
\end{equation}
where $T$ is the temperature, $k_B$ is the Boltzmann constant, $\beta=1/k_BT $ and $p({\boldsymbol{s}})$ is a target distribution which is to be normalized. 
It has been shown \cite{Valsson2014} that $\Omega[V({\boldsymbol{s}})]$ is a convex functional and at its minimum the value for the bias $V( {\boldsymbol{s}} )$ is, within an irrelevant constant, 
\begin{equation}
\label{bias}
V( {\boldsymbol{s}} ) = - F({\boldsymbol{s}}) - \frac{1}{\beta} \log p({\boldsymbol{s}}) \, .
\end{equation}
%add C
From equation \eqref{bias} it follows that at the minimum the CV distribution in the biased ensemble, 
\begin{equation}
\label{pv}
P_V( {\boldsymbol{s}} ) = \frac{e^{-\beta (V({\boldsymbol{s}}) + F({\boldsymbol{s}}))}}{\int d{\boldsymbol{s}} \ e^{-\beta (V({\boldsymbol{s}}) + F({\boldsymbol{s}}))} } \, , 
\end{equation}
is equal to the pre-assigned target distribution $p({\boldsymbol{s}})$. %This is in fact one of the strengths of VES.
\paragraph*{Optimization}
In order to use the variational principle and minimize the functional, the bias $V( {\boldsymbol{s}} )$ is expressed in terms of a set of parameters ${ \boldsymbol{\alpha} } = \{ \alpha_1, \alpha_2, \alpha_3, ... \}$.  This turns the functional $\Omega[V({\boldsymbol{\alpha}})]$ into a function $\Omega({\boldsymbol{\alpha}})$. This function $\Omega({\boldsymbol{\alpha}})$ is then optimized using the gradient  
\begin{equation}
\label{gradient}
\frac{\partial \Omega( {\boldsymbol{\alpha}} ) }{\partial \alpha_i}  =  - \ \Big\langle { \frac{\partial V( {\boldsymbol{s}; \boldsymbol{\alpha}} ) }{\partial \alpha_i} }\Big\rangle _{V( {\boldsymbol{\alpha}} )} \ + \ \Big\langle { \frac{\partial V( {\boldsymbol{s}; \boldsymbol{\alpha}} ) }{\partial \alpha_i} }\Big\rangle _{p} \, ,
\end{equation}
where the terms $\langle \cdots \rangle _{V( {\boldsymbol{\alpha}})}$ and $\langle \cdots \rangle _{p}$ are expectation values calculated in the ensemble biased by ${V( {\boldsymbol{\alpha}})}$ and over the distribution $p({\boldsymbol{s}})$ respectively. However, the gradient evaluated from such a statistical calculation is inherently noisy and thus, a stochastic optimization algorithm \cite{stochastic} is necessary to obtain a faster and smoother convergence. 

Here, we take the bias potential $V( {\boldsymbol{s}}; {\boldsymbol{\alpha}} ) $ to be a linear expansion in a finite number of basis functions, 
\begin{equation}
\label{Vlinexp}
V( {\boldsymbol{s}}; {\boldsymbol{\alpha}} ) = \sum_k \alpha_k f_k(\boldsymbol{s}) \, , 
\end{equation}
and use the Bach and Moulines averaged stochastic gradient descent algorithm \cite{BachMoulines} to update iteratively the parameters of the bias expansion. 

\paragraph*{Target distribution}
A straightforward choice of $p({\boldsymbol{s}})$ is the uniform distribution, that at convergence gives $V({\boldsymbol{s}})= -F({\boldsymbol{s}})$. However, this distribution forces the system to visit with equal probability all points in the CV space, including uninteresting high energy regions, often resulting in poor convergence. Although it is possible to construct a $p({\boldsymbol{s}})$ that focuses on important regions of the CV space, this requires apriori information on $F({\boldsymbol{s}})$, a quantity that is in general unknown at the beginning of the calculation. 
However, \citeauthor{Valsson2015} \cite{Valsson2015} introduced a self-consistent approach to include $F({\boldsymbol{s}})$ in the definition of $p({\boldsymbol{s}})$. They suggested a $p({\boldsymbol{s}})$ that at convergence results in,
\begin{equation}
\label{WTDves}
p({\boldsymbol{s}})= \frac{e^{-{\frac{\beta}{\gamma}} F({\boldsymbol{s}})}}{\int{ d{\boldsymbol{s}} \  e^{-{\frac{\beta}{\gamma}} F({\boldsymbol{s}})}} } \, ,
\end{equation}
where $\gamma \geq 1 $ is a parameter called bias factor. Unlike VES with a fixed $p({\boldsymbol{s}})$, this approach requires two coupled iterative schemes.  In addition to updating the parameters for $V( {\boldsymbol{s}}; {\boldsymbol{\alpha}} ) $ every $\Delta t_{\text{V}}$ MD steps, the $p({\boldsymbol{s}})$ is updated every $\Delta t_{\text{TD}}$ number of iterations of the optimization of $\Omega( {\boldsymbol{\alpha}} )$. At first, $p^{(0)}({\boldsymbol{s}})$ is typically taken to be a uniform distribution and  the first guess for $F({\boldsymbol{s}})$ is zero. The $F({\boldsymbol{s}})$ at any given step $(k+1)$ is then obtained as,
\begin{equation}
\label{self-consistent}
\begin{split}
F^{(k+1)}( {\boldsymbol{s}} ) &= - V^{(k)}({\boldsymbol{s}}) - \frac{1}{\beta} \ \log p^{(k)}({\boldsymbol{s}}) \\
&= - V^{(k)}({\boldsymbol{s}}) - \frac{1}{\gamma} \ F^{(k)}({\boldsymbol{s}}) \, ,
\end{split}
\end{equation}
In the asymptotic limit, one has $V({\boldsymbol{s}}) = - ( 1 - \frac{1}{\gamma} ) \ F({\boldsymbol{s}})$ modulo a constant as in the well-tempered ensemble.

\begin{figure}[h]
\centering 
 \includegraphics[scale=0.6]{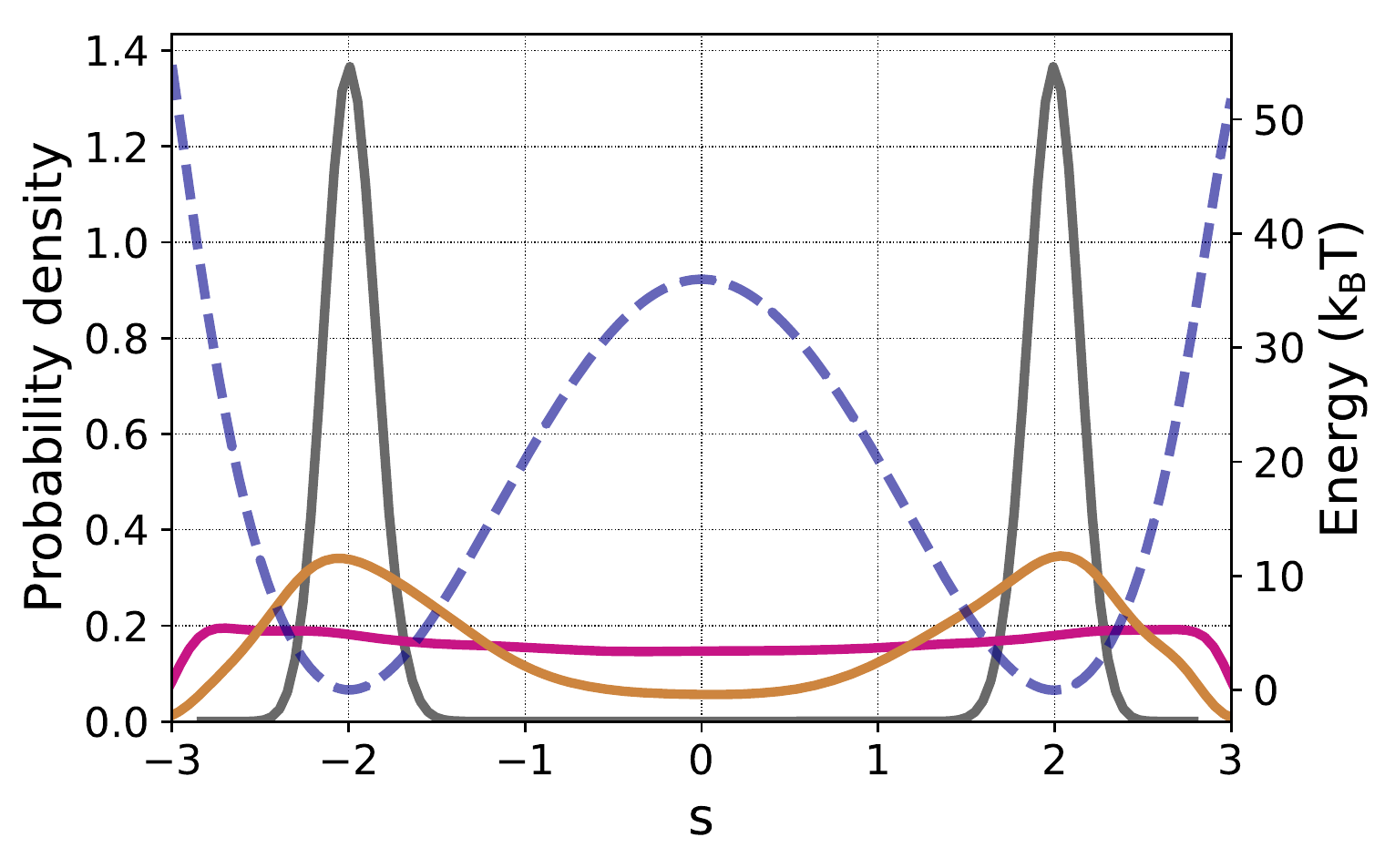}
 \caption[]%
 {{\small A comparison of the biased and unbiased s probability densities obtained using uniform (magenta) and well-tempered (yellow) target distributions for a one-dimensional double well potential $U(x) = 2.25 x^4 - 18 x^2 + 36$. The free energy (blue, dashed lines) obtained using the well-tempered simulation and the unbiased probability density (grey) are marked for reference. The probability densities were obtained using gaussian kernels. All simulation details are included in the Supplementary Information. In the unbiased case, we performed separate calculations for each different minimum energy state.}}    
 \label{fig:unbiased}
\end{figure}

It has been shown that, even if the instantaneous estimate of $F({\boldsymbol{s}})$ is initially inaccurate, VES quickly converges to the correct $F({\boldsymbol{s}})$ \cite{Valsson2014,Valsson2015}. Additionally, in many practical cases, the well-tempered variant was found to converge much faster than a VES simulation that uses a uniform $p({\boldsymbol{s}})$. In figure \ref{fig:unbiased}, we illustrate with the case of a one-dimensional double-well potential of barrier height $40 \ k_B T$, the probability density of the CV obtained with uniform and well-tempered $p({\boldsymbol{s}})$ as compared to the unbiased Boltzmann probability distribution. It can be seen that in a well-tempered simulation, the system visits the TS less frequently than in a uniform target simulation. However, as discussed above one has to pay the price of a slower convergence. 

\subsection{Targeting the transition state}
For simplicity sake, we deal here only with cases in which the CV is unidimensional. However, an extension to a multidimensional set of CVs is possible.  For a well chosen CV, the TS is located at the apparent $F(s)$ TS. Hence, sampling of the TS region can be enhanced by designing a $p({{s}})$ that gives a higher probability of sampling in a region around the apparent TS. 

To do so, we utilize the instantaneous $F({{s}})$ similarly to what is done in well-tempered VES. We take into account the fact that the apparent TS is a $F({{s}})$ stationary point, where $\frac{dF({s})}{ds}=0$. A $p(s)$ constructed to enhance the probability of sampling the stationary points should, in principle, be able to boost TS sampling. The most obvious choice of such a distribution would be a Gaussian-like distribution similar to the well-tempered $p(s)$ that has as argument $\frac{dF({s})}{ds}$. However, we found it difficult to describe the rapidly varying tail of such a function with a finite basis expansion unless $\gamma$ was taken so large that the resulting distribution was practically uniform. Hence, we ameliorated the behaviour at the tail by using a Lorentzian-like function that decays polynomially
\begin{equation}
\label{lorgrad}
p_L({s})= \frac{1}{Z_p} \ \frac{\zeta}{\zeta^2 +  \left({{\beta \frac{d F({s})}{d s} }}\right)^2} \, ,
\end{equation}
where $Z_p$ is a normalization factor, and $\zeta$ is a parameter of units $s^{-1}$ that regulates the broadening of the distribution. 
This $p_L(s)$ will weight uniformly all the stationary states of $F(s)$: minima, TS, and if present, inflection points. However for the purpose of our calculation, it is desirable to give more weight to the TS. In order to do this we take advantage of the fact that at a local maximum not only $\frac{dF({s})}{ds}$ vanishes but also the second derivative satisfies the condition $\frac{d^2F({s})}{ds^2} < 0$. For this reason, 
we multiply $p_L({s})$ in equation \eqref{lorgrad} by the function,
\begin{equation}
\label{switch}
S_{a,\lambda}(s)=   \ \frac{1-a}{1 \ + \ e^{ \lambda \beta \frac{d^{2}F({s})}{ds^2}}} + a \, ,
\end{equation}
where $0 \leq a \leq 1$. This function goes from the value $S_{a,\lambda}(s)=1$, when $\frac{d^2F({s})}{ds^2}$ is large and negative to $S_{a,\lambda}(s)=a$, when $\frac{d^2F({s})}{ds^2}$ is large and positive. The parameter $\lambda$ determines the sharpness of this transition. In practice for $\lambda \to \infty $ and $a=0$, $S_{a,\lambda}(s)$ behaves like a rectangular pulse and has a non-zero value in the region in which $\frac{d^2F({s})}{ds^2} \leq 0$. This region from now on is taken to define the TS region. The case $a=1$ is included for generality and corresponds to using simply $p_L(s)$. This can come useful when an equilibrated sampling between maxima and minima is required. Thus, the resulting $p(s)$ is,
\begin{equation}
\label{modsw}
\begin{split}
p({s}) &= \frac{1}{Z_p} \ \frac{\zeta}{\zeta^2 +  \left({\beta \frac{d F({s})}{d s}}\right)^2}\left( \frac{1-a}{1 \ + \ e^{ \lambda \beta \frac{d^{2}F({s})}{ds^2}}} + a\right) \\
&= p_L(s) S_{a,\lambda}(s) \, .
\end{split}
\end{equation}  

\begin{figure}[h]
\centering 
 \includegraphics[scale=0.6]{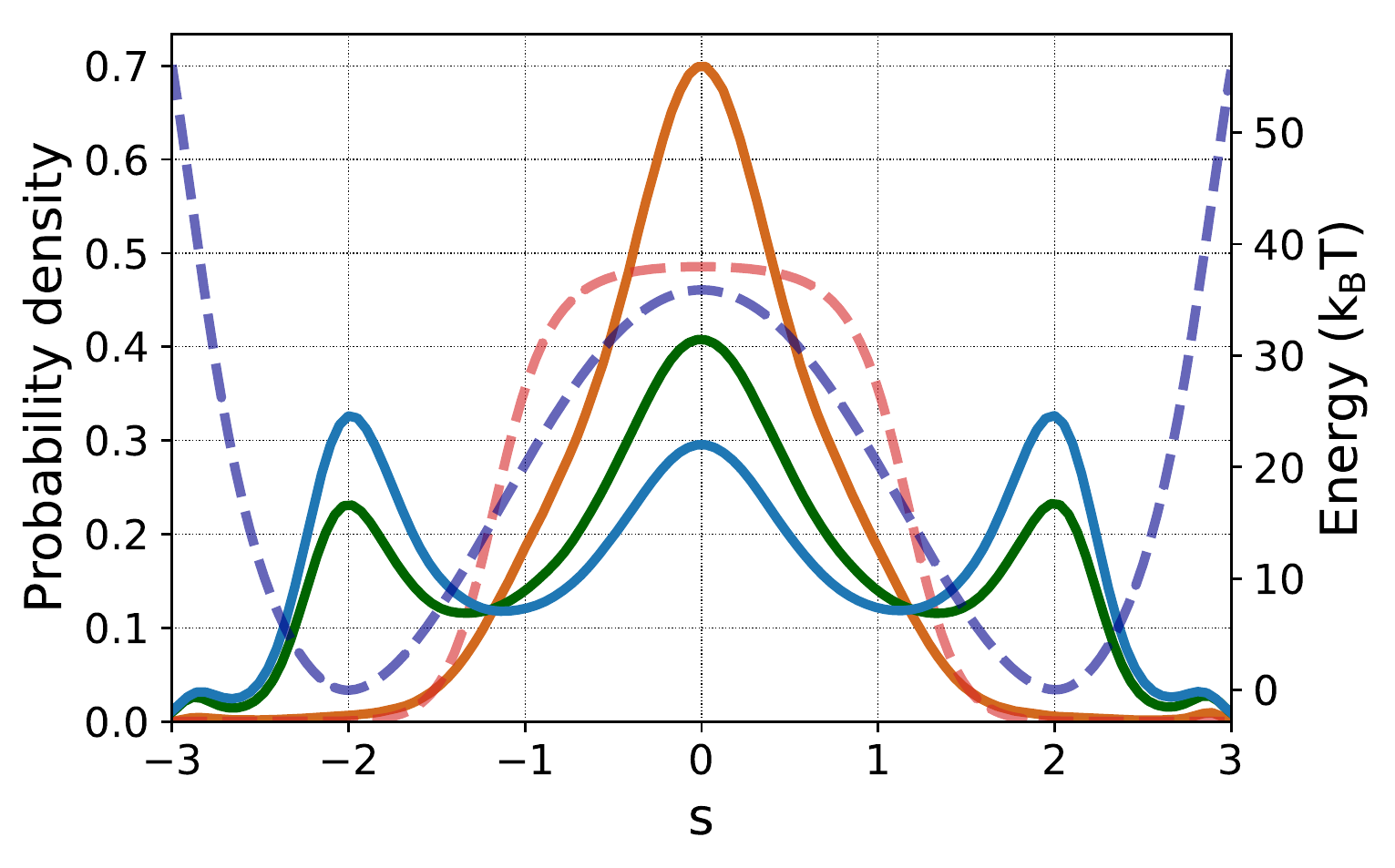}
 \caption[]%
 {{\small The working principle of the target distribution is shown for the symmetric one-dimensional double well potential. The parameter $a$ is used to modulate the probability in the different regions. Three different target distributions are compared: $p_L(s)$ with $\zeta = 20$ in blue, $p_L(s) \ S_{0.5,0.1}(s)$ in green and $p_L(s) \ S_{0.01,0.1}(s)$ in orange. The free energy (blue, dashed lines) and $S_{0.5,0.1}(s)$ (red, dashed lines) are marked for reference ($S_{0.5,0.1}(s)$ has been shifted by $0.5$ along the y-axis for clarity). }}    
 \label{fig:all}
\end{figure}

In figure \ref{fig:all}, we show the effect of changing the parameters $a$ and $\lambda$ in the symmetric double well case. 

In order to obtain these results and the others reported below we had to solve the problem of computing the first and second derivatives of $F(s)$. If not done with care this can introduce unwanted noise. To this effect we expand $F({s})$ 
in the same basis set $\{f_k({s})\}$ used to express $V(s)$ (see equation \eqref{Vlinexp}). In this way, the $s$ dependence appears only in the functions $f_k(s)$ whose derivatives can be computed analytically. 

We now illustrate how this approach works in practice on the paradigmatic cases of a bimolecular nucleophilic substitution reaction and the condensation of a Lennard-Jones gas.

\section{Bimolecular nucleophilic substitution reaction}
Nucleophilic substitution reactions constitute a popular class of chemical reactions. The asymmetric substitution reaction of fluoromethane and chloromethane CH${_{\text{3}}}$F + Cl${^{\text{-}}} \ {\rightleftharpoons} \ $ CH${_{\text{3}}}$Cl + F${^{\text{-}}}$ is a simple example of such reactions and has been well studied over the years. Following a previous enhanced sampling study of the same reaction \cite{Piccini2017}, we chose the difference between the C-Cl ($d_{Cl}$) and C-F ($d_F$) atomic distances as  CV. 

Here, we contrast four different simulations that use VES with different target distributions. The first is a well-tempered one with $\gamma=50$, the second uses a $p(s)$ that is uniform in the interval $-2.0 \leq s \leq 2.9$. The third simulation uses $p_L(s)$ as defined in equation \eqref{lorgrad} with $\zeta=20$ {\AA}$^{-1}$. 
Finally, in the last one we use $p_L(s)$ given by $\zeta=40$ {\AA}$^{-1}$ and multiplied by a $S_{a, \lambda}(s)$ that has $a=0.01$ and $\lambda=0.1$. In the last two cases the values of the parameters have been chosen such that the resulting $p(s)$ does not present very sharp features, thus limiting the number of basis functions that need to be included in the $V(s)$ expansion. Our choice of basis functions was to use orthonormal Legendre polynomials. A detailed description of the VES parameters and the simulation details is provided in the supplementary information (SI).    
\begin{figure}[h]
\centering
\includegraphics[scale=0.6]{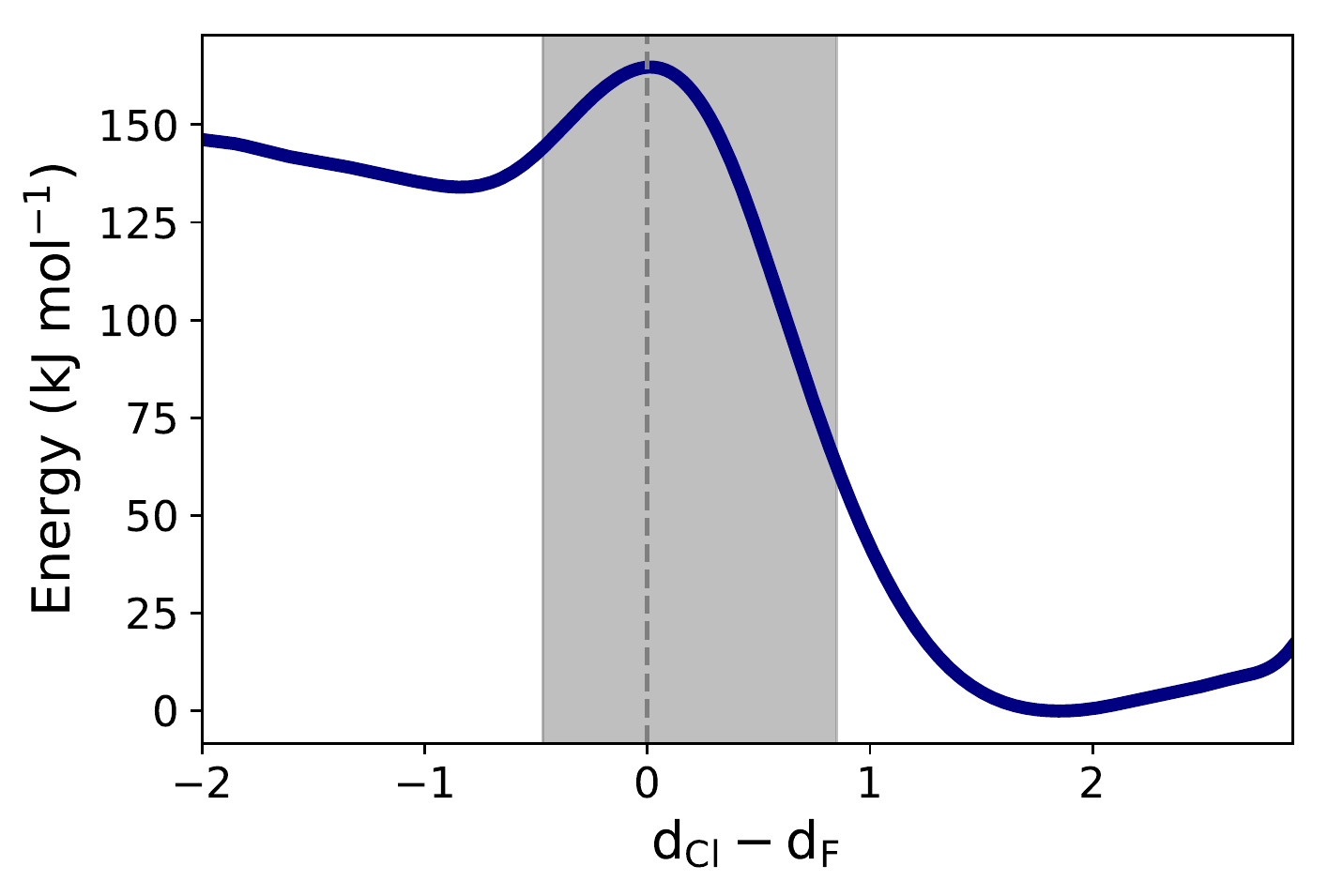}
\caption{Reweighted free energy obtained for nucleophilic substitution reaction of fluoromethane and chloromethane. The shaded region corresponds to the TS region as defined in the main text.   \label{fig:sn2fes} }
\end{figure}

We first converge the bias coefficients (${\boldsymbol{\alpha}}$) and use the resulting static potential to perform the simulation. We then obtain the  $F(s)$ using an umbrella-like reweighting. As expected all the four calculations give the same $F(s)$ within statistical errors. In table \ref{tab:recrossings}, we quantify the fraction of configurations in the TS region and the number of times the system makes a transition between the two minima. One can observe that the VES calculation with the $p_L(s) \ S_{a, \lambda}(s)$ target distribution has indeed resulted in a larger number of configurations sampled in the TS region. Furthermore, it can be seen from the table that the number of transitions is greatly enhanced. 
\begin{table}[htb]
\begin{center}
  \begin{tabular}{ | l | c | c |}
    \hline
    \textbf{Simulation} & \textbf{Configurations in TS region (\%)} & \textbf{Number of recrossings} \\ \hline
    Well tempered & 19.02 & 229 \\ \hline
    Uniform & 23.19 & 253 \\ \hline
    $p_L(s)$ & 28.38 & 233\\ \hline
    $p(s)$ & 63.68 & 339 \\ 
    \hline
  \end{tabular}
\end{center}
\caption {Percentage of configurations observed in the TS region and the number of transitions between the two minimum energy states for 500 ps long simulations.  } \label{tab:recrossings}
\end{table}

\section{Liquid-vapour phase transition}
Nucleation is a phenomena of interest in a broad range of fields. A commonly used model in homogeneous nucleation studies is that of the Classical Nucleation Theory (CNT). It states that the energetics of growth of a stable nucleus (or cluster) in a metastable bulk is shaped by an interplay of two factors. While the creation of the interface between this cluster and the bulk phase hinders the cluster growth, the formation of these more stable regions favours it. These two terms are proportional to the surface area and the volume of the cluster respectively. As a result of these competing contributions, cluster growth is hindered until a critical size is reached. Beyond this critical size, the stability gained from the volume term surpasses the cost of interface formation and the cluster grows spontaneously. Due to the existence of this barrier, it is often difficult to determine the critical cluster size \cite{Sun2018}\cite{Auer2001} and obtain sufficient sampling in the region even using enhanced sampling techniques.

\begin{figure}[]
    \includegraphics[scale=0.6]{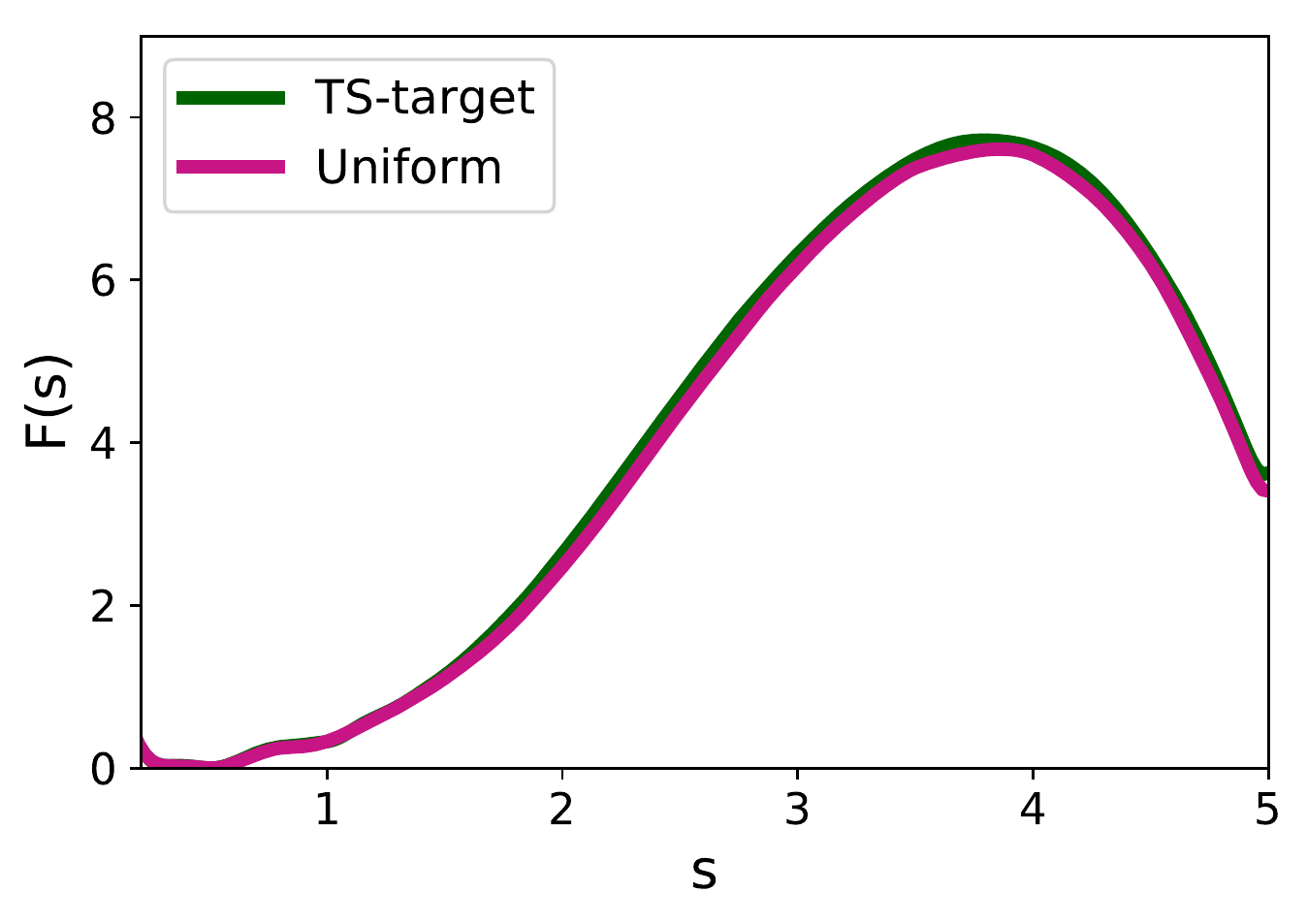}
	\caption{The reweighted free energy surface obtained from the two simulations for the condensation of a Lennard-Jones gas. The free energy surface obtained from both the simulations are equivalent within statistical errors. 
	\label{fig:ljfes}}
\end{figure}

Here, we compare the results of our TS-target VES simulation with uniform VES simulation in the case of condensation of Argon. The details of the simulation can be found in the SI.
Following \citeauthor{Piaggi2016b}\cite{Piaggi2016b}, we use a CV defined as $s=(n_l)^{1/3}$ where $n_l$ is the number of liquid-like atoms. In the case in which there is only one cluster present $s$ is roughly proportional to the radius of the cluster. We further restrict the growth of clusters beyond $120$ atoms using a harmonic restraining potential having $\kappa = 0.1$. The parameters for the uniform VES simulation are similar to those of \citeauthor{Piaggi2016b} \cite{Piaggi2016b} while that of the TS-target are $a=0.02$, $\lambda = 0.1$ and $\zeta=30$. Since we want to focus predominantly in the critical region we use a very small $a$. Here also, we first converge the bias coefficients $( {\boldsymbol{\alpha}} )$ for the two cases using $10$ walker simulations and then perform a single walker simulation with a static bias potential. The $F(s)$ is obtained by an umbrella-like reweighting and is shown in figure \ref{fig:ljfes}. 

\begin{figure}[]
\centering
\includegraphics[scale=0.3]{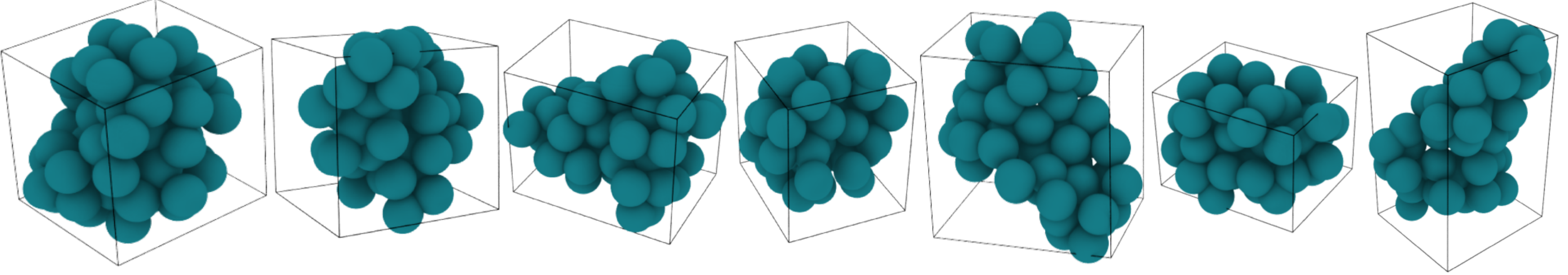}
\caption{Snapshots of clusters (54 atoms) having different shape anisotropy (k).  \label{kvalues} }
\end{figure}

\begin{figure}[]
	\includegraphics[scale=0.9]{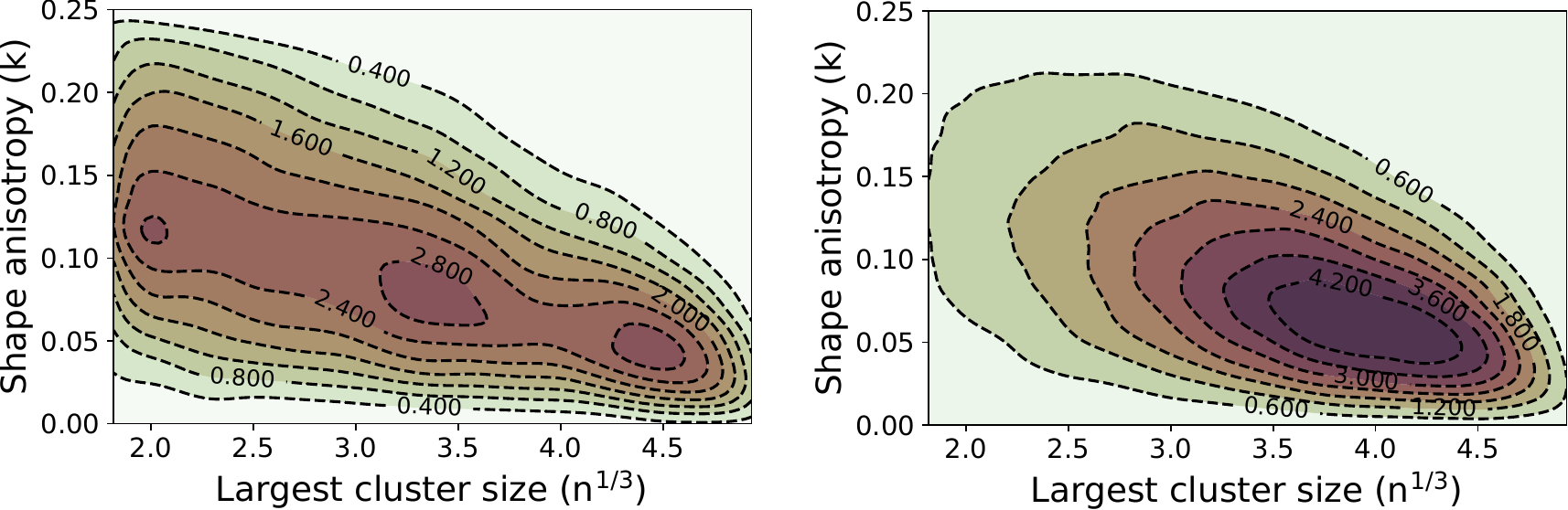}
	\caption{Probability density of the collective variable space sampled expressed as a function of the shape anisotropy value ($k$) and the size of the largest cluster ($s=n^{1/3}$) obtained with the two target distributions ( left:Uniform, right:TS-target ). The probability density values of the different regions are marked using contours. With the TS-target, the shape of the critical cluster is thus estimated more accurately. Only clusters having at least 10 atoms have been considered in the plots
         \label{fig:kvalue}}    
\end{figure}

Further, we analyze the shapes of the largest clusters obtained during the simulations. To do so, first we identify the largest clusters of liquid-like atoms obtained from both the simulations using the depth first search clustering algorithm described in Ref.~\citenum{Tribello2017}. Clusters spanning more than half the box size and clusters with less than 6 atoms are rejected. We calculate the moment of inertia tensor matrix, ${\boldsymbol{T}}$ for all the clusters using $T_{\alpha\beta} = \sum ' m( x_{\alpha \beta} \ - \ \hat{x}_{\alpha \beta} )$, where the primed sum is over the atoms contained in the largest cluster, $m$ and $x$ are the mass and position of these atoms while $\hat{x}$ is the position of the center of mass of the cluster. We diagonalize the matrix to obtain the three eigenvalues $( \lambda_1 \geq \lambda_2 \geq \lambda_3 )$ and subsequently calculate the shape anisotropy value defined as $k = \frac{3}{2} \frac{ \lambda_1^4 \ + \ \lambda_2^4 \ + \ \lambda_3^4}{(\lambda_1^2 \ + \ \lambda_2^2 \ + \ \lambda_3^2)^2} -\frac{1}{2} $. The value $k=0$ indicates a perfectly spherical cluster whereas the value $k=1$ corresponds to a cluster where all the particles are linearly arranged. Figure \ref{kvalues} shows how even clusters with the same number of particles could have very different shapes and hence, a range of anisotropy values.  In figure \ref{fig:kvalue}, the probability density of the CV space is shown. The probability density has been estimated using multivariate gaussian kernels.

It can be concluded from figure \ref{fig:kvalue} that the proposed target distribution has been able to sample more effectively the critical cluster size. With the uniform target distribution, the system tends to sample more the uninteresting region with very small clusters due to the low energy of the region while with the TS-target distribution, the high-energy region around the critical cluster has been sampled more in the same computational time. 
Also, it is evident from both the figures that as the cluster size increases the clusters tend to be more spherical. Hence, while the results obtained conform to previously observed trends, we are now able to quantify more confidently about the critical cluster properties. 

\begin{figure}[h]
	\centering
	\includegraphics[scale=0.6]{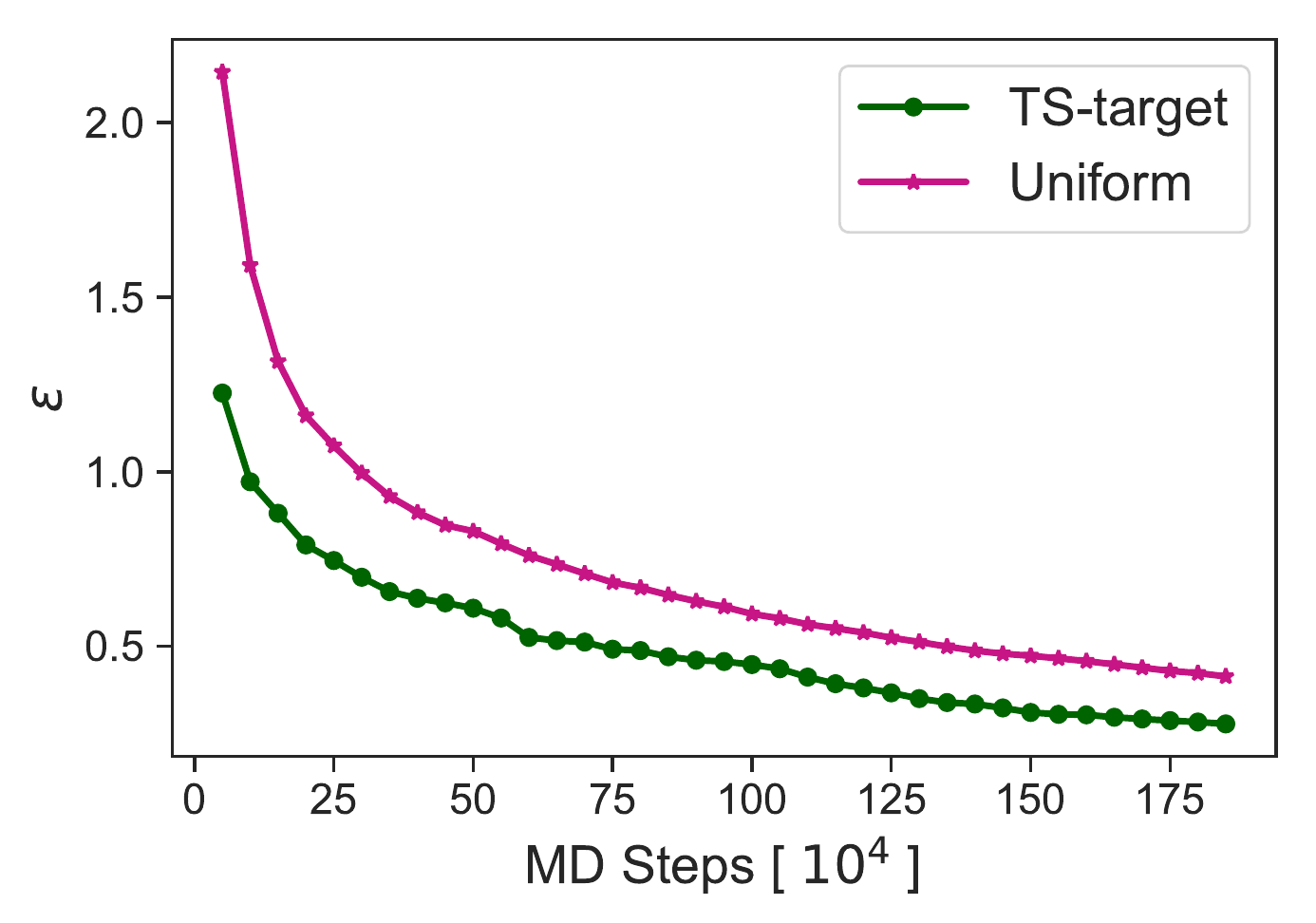}
	\caption{The error ($\varepsilon$) calculated using Eq. \eqref{error}. Multiple $F(s)$ were obtained by reweighting 10 walker simulations at intervals of $5 \times 10^4$ MD steps. The reference $F(n^{1/3})$ was obtained by reweighting a simulation using the converged bias from the 10 Walker Uniform VES simulation.
	\label{fig:ljconv}}
\end{figure}

Last but not least in this case an improvement in convergence rate is observed. In order to quantify the convergence of $F(s)$, we use an error metric similar to that used by \citeauthor{Valsson2015} \cite{Valsson2015}. 
\begin{equation}
\label{error}
\varepsilon = \sqrt{ \frac{\int d {{s}} \ [\bar{F}_r({{s}}) - \bar{F}({{s}})] }{\int d {{s} }  }}
\end{equation}
where 
\begin{equation}
\label{error_r}
\bar{F}_r({{s}}) = F_r({{s}}) - \frac{\int d {{s} } \ F_r({{s}})  }{\int d {{s} }  }
\end{equation}
\begin{equation}
\label{error_s}
\bar{F}({{s}}) = F({{s}}) - \frac{\int d {{s} } \ F({{s}})  }{\int d {{s} } }
\end{equation}
Multiple $F(s)$ are obtained by reweighting the Uniform and TS-target 10 walkers VES simulations at regular intervals using the reweighting scheme proposed in Ref.~\citenum{Yang2018}. The reference $F(s)$ is obtained by an umbrella-like reweighting of the simulation that used the converged static bias obtained from the 10 walker Uniform VES simulation. It can be seen from figure \ref{fig:ljconv} that the TS-target does converges faster. 

\section{Conclusion}
We have proposed a very general approach based on the Variationally Enhanced Sampling method to enhance sampling in the transition region, and exemplified with the cases of a S$_\text{N}$2 reaction and condensation of Lennard-Jones gas how our method can be successfully applied to very diverse scenarios. Furthermore, our approach is capable of locating the transition states on the fly and acquiring an increased sampling of the transition region despite the process being a rare-event. This of course under the condition of dealing with a good quality CV. Taking into account the ability of our method to increase the frequency with which the transition region is visited, we believe that this could be especially useful in studying reactions that take place in condensed phases where the effect of the environment on the transition state is relevant and yet difficult to describe with a single conformation. Nucleation is one example of such class of reactions. We believe our approach in these cases may obtain a much faster $F(s)$ exploration. 
\begin{acknowledgement}

The authors would like to thank Tarak Karmakar for carefully reading the manuscript and for useful discussions. The authors thank the European Union Grant No. ERC-2014-ADG-670227 and the National Center for Computational Design and Discovery of Novel Materials MARVEL for funding. The calculations were performed using the Euler HPC Cluster at ETH Zurich and the M\"onch cluster of the Swiss National Computing Center (CSCS).

\end{acknowledgement}

\begin{suppinfo}

The computational details for the one-dimensional potential, the S$_\text{N}$2 reaction and the condensation of Argon are included as Supporting Information. 

\end{suppinfo}

\bibliography{Enhancing-TS-arxiv}
\end{document}